\documentclass[conference]{IEEEtran}

\usepackage{graphicx}  
\usepackage{amsmath, amssymb}  
\usepackage{cite}  
\usepackage{etoolbox}  
\usepackage{enumitem}
\usepackage{subcaption}
\usepackage{url}
\usepackage{hyperref}

\makeatletter
\patchcmd{\section}{\noindent}{}{}{}
\patchcmd{\subsection}{\noindent}{}{}{}
\patchcmd{\subsubsection}{\noindent}{}{}{}
\patchcmd{\@sect}{\@afterindenttrue}{\@afterindentfalse}{}{}
\makeatother

\begin{document}

\title{ Unmasking COVID-19 Vulnerability in Nigeria: Mapping Risks Beyond Urban Hotspots
}

\author{ \textbf{TEAM MEMBERS:}\\
    \IEEEauthorblockN{Sheila Wafula, Blessed Madukoma}  
    \IEEEauthorblockA{
        Carnegie Mellon University, Africa \\  
        Kigali, Rwanda \\
        \{swafula, bmadukom\}@andrew.cmu.edu
    }
}

\maketitle

\begin{abstract}
The COVID-19 pandemic has presented significant challenges in Nigeria's public health systems since the first case reported on February 27, 2020. This study investigated key factors that contribute to state vulnerability, quantifying them through a composite risk score integrating population density (weighted 0.2), poverty (0.4), access to healthcare (0.3), and age risk (0.1), adjusted by normalized case rates per 100,000. The states were categorized into low-, medium-, and high-density areas to analyze trends and identify hotspots using geographic information system (GIS) mapping. The findings revealed that high-density urban areas, such as Lagos, which account for 35.4\% of national cases, had the highest risk scores (e.g., Lagos: 673.47 vs. national average: 28.16). These results align with global and local studies on the spatial variability of COVID-19 in Nigeria, including international frameworks such as the CDC Social Vulnerability Index. Google Trends data highlighted variations in public health awareness, serving as a supplementary analysis to contextualize vulnerability. The risk score provides a prioritization tool for policy makers to allocate testing, vaccines, and healthcare resources to high-risk areas, although data gaps and rural underreporting call for further research. This framework can extend to other infectious diseases, offering lessons for future pandemics in resource-limited settings.


\end{abstract}

\maketitle

\maketitle
\section{Introduction}
The COVID-19 pandemic presented significant public health challenges around the world since the World Health Organization declared it a pandemic on 11 March 2020.
Its impact has varied widely across countries and regions, highlighting the importance of factors such as population characteristics, economic conditions, healthcare systems, and government policies \cite{Jonah2021PublicHealth}. These differences show why targeted, data-driven approaches are needed to assess vulnerability and manage the pandemic effectively.

In Nigeria, the pandemic presented complex challenges due to the diverse population and socioeconomic conditions of the country. The first confirmed case was reported on 27 February 2020 in Lagos, involving an Italian citizen. This marked the beginning of varied transmission patterns across Nigerian states \cite{Owhonda2021COVID19}. 

Similarly to global trends, urban areas, especially Lagos, experienced a higher burden, accounting for 35.4\% of national cases, with much higher infection rates than rural regions \cite{Jonah2021PublicHealth}. At the time of this study, Nigeria had more than 260,000 confirmed cases and a fatality rate of approximately 1.18\% \cite{Taye2021}.

To improve response to pandemics, tools that combine multiple factors to assess vulnerability are critical. The COVIRA (COVID-19 Vulnerability and Risk Assessment) framework, originally developed for Nepal, is one of these tools \cite{Parajuli2020}. It calculates risk by considering COVID-19 cases, health conditions, healthcare availability, economic factors, and immigration data, producing a score from 0 to 100 \cite{Parajuli2020}. Since Nigeria shares similarities with Nepal, such as being a lower-middle-income country with significant urban-rural disparities and limited testing capacity, COVIRA is a relevant model for Nigeria \cite{Taye2021}. This makes it a valuable starting point to understand vulnerability in the unique context of Nigeria.

This study addresses the research question: \textbf{What are the primary factors that contribute to COVID-19 vulnerability in Nigerian states and how can they be quantified using a composite risk score to inform decision-making?} Nigeria’s demographic and socioeconomic diversity creates uneven risks, which requires targeted public health strategies \cite{Taye2021}. Unlike previous studies that focus on single factors such as population density or poverty \cite{Jonah2021PublicHealth}, \cite{Hassan2020COVID19}, this research integrates multiple dimensions into a unified vulnerability score.

The research develops a composite risk score to quantify state-level vulnerability in Nigeria, adapting the COVIRA framework \cite{Parajuli2020}. The score incorporates four weighted factors, adjusted by normalized case rates per 100,000 to reflect current vulnerability:
\begin{enumerate}
    \item Population density \texttt{(weight: 0.2)} - which reflects the transmission potential in crowded areas.
    \item Poverty (\texttt{weight: 0.4)} - assigned the highest weight due to its impact on healthcare access and living conditions, especially in rural areas with potential under-reporting.
    \item Access to healthcare \texttt{(weight: 0.3)} - accounts for uneven facility distribution across states.
    \item Age risk \texttt{(weight: 0.1)} - given the lowest weight, as Nigeria's young population mitigates severe outcomes.

\end{enumerate}

Geographic Information System (GIS) mapping classifies states into low-, medium- and high-risk groups, improving visualization of case distributions and hotspots.

Preliminary findings indicate that high-density urban areas, particularly Lagos, are disproportionately affected. Lagos recorded a risk score of 673.47, far exceeding the national average of 28.16, and accounted for 35.4\% of cases, consistent with prior spatial analyses \cite{Hassan2020COVID19}. Beyond density, poverty and limited access to healthcare increase risk, especially in rural areas with socioeconomic challenges despite lower population densities. Rural underreporting, due to inadequate testing and data collection, suggests that actual vulnerability may be higher than scores indicate. Poverty and access to healthcare remain critical drivers in both urban and rural settings.

This study develops a composite risk score to quantify the vulnerability of COVID-19 in Nigerian states, integrating population density, poverty, access to healthcare, and age risk, with supplementary Google Trends data to contextualize public health awareness (e.g., declining interest after March 2020). This data-driven framework, adapted from the COVIRA model for the context of Nigeria, prioritizes the allocation of resources for testing, vaccines and healthcare in high-density urban and socioeconomically vulnerable rural areas. 

This approach addresses the diverse epidemiological landscape of Nigeria, offering a robust tool for public
health interventions.  These findings highlight
the potential for data-driven strategies to improve pandemic
response and inform public health policy in Nigeria. Although focused on COVID-19, the framework can extend to other infectious diseases, offering lessons for future pandemics in resource-limited settings.

\section{Literature Review}
\subsubsection{Global Frameworks and Approaches}
Across the world, other studies have developed similar ways of measuring risk by combining many different factors. This validates that multidimensional vulnerability assessment is a good approach to understanding epidemics.
In India, researchers created a detailed vulnerability index based on 15 factors in five main areas: socioeconomic status, demographic composition, housing and hygiene conditions, epidemiological factors, and capacity of the health system. This method successfully identified highly vulnerable areas in nine large states and helped leaders decide where to allocate resources during the pandemic. The study noted that their vulnerability index aligned with where COVID-19 cases were concentrated at the state level \cite{Acharya2020}.

In Italy, a risk model was focused on three key parts: disease hazard, area exposure, and vulnerability of its people. This approach explained why northern regions such as Lombardia, Emilia-Romagna, Piemonte, and Veneto suffered more from COVID-19. The study revealed a strong connection between the predicted risk classification and the actual impacts, including infection rates, intensive care needs, and deaths \cite{Pluchino2021}.

In Africa, Kenya developed three vulnerability indices at the sub-county level: a Social Vulnerability Index (SVI) with 19 indicators on socioeconomic deprivation, access to services and population dynamics; an Epidemiological Vulnerability Index (EVI) with five factors on comorbidities; and a combined Social-Epidemiological Vulnerability Index (SEVI). The SEVI explored the general resilience and risk of severe COVID-19, grouping subcounties into vulnerability ranks \cite{Macharia2020}.

South Africa created the COVID-19 Vulnerability Index (SACVI) using 2011 Census data, finding that 40\% of the population was vulnerable, mainly due to poor sanitation and household crowding. The SACVI aimed to identify populations at increased risk of SARS-CoV-2 infection and severe outcomes \cite{StatsSA2020}.

Globally, the US CDC's Social Vulnerability Index (SVI) was adapted for COVID-19 to target resources to vulnerable communities, demonstrating its utility in pandemic responses \cite{Chien2024}.

\subsubsection{Nigerian Context and Research Gaps}
Earlier research on COVID-19 in Nigeria primarily examined individual risk factors rather than a broad, combined approach. For example, Taye and Popoola analyzed spatial variability, showing that population density was significant for transmission \cite{Taye2021}. Their work highlighted the differences between urban and rural areas, with Lagos, the Federal Capital Territory (FCT), the Plateau, and Rivers accounting for many cases.

Although Adams and Obaroni modeled density alongside socioeconomic and environmental factors, they did not combine these variables into a unified score \cite{Adams2023PopulationDensity}. This gap is critical, as densely populated areas are more susceptible to transmission, and limited access to healthcare increases vulnerability, particularly in rural regions with high poverty \cite{Taye2021}.

Furthermore, Hassan and Hashim noted the challenge of rural under-reporting of COVID-19 cases\cite{Hassan2020COVID19}. Adesola et al. noted healthcare disparities, supporting rural-focused interventions \cite{Adesola2024PopulationGrowth}. Nigeria’s diverse infrastructure and information consumption patterns require a comprehensive assessment accounting for interacting factors \cite{Adesola2024PopulationGrowth}.

\subsubsection{Framework Justification}
The lack of a comprehensive vulnerability assessment for Nigeria is especially crucial given the complex demographic and socioeconomic landscape of the country \cite{Usman2024}. Although international frameworks demonstrate multifactor risk measurement approaches, their application to Nigeria may require adjustments to fit local conditions, potentially including customized weighting schemes that reflect the country's unique vulnerability patterns.

This study fills this gap by developing a composite risk score for Nigerian states, which combines population density, poverty, access to healthcare, and age risk, adjusted for normalized case rates.
The approach builds on the successful COVIRA framework while incorporating lessons learned from other international implementations. 

This enables public health authorities to prioritize interventions in both high-risk urban areas and vulnerable rural regions \cite{Hassan2020COVID19}. The research informs resource allocation and policy decisions, improving Nigeria's ability to effectively manage current and future health emergencies.

\section{Methodology}
\subsection{Overview}
This study evaluates the vulnerability of COVID-19 in Nigerian states by developing a composite risk score, visualizing spatial patterns with geographic information system (GIS) mapping, and applying statistical analysis to identify key factors and guide public health strategies. The methodology covers data preparation, exploratory analysis, risk score construction, spatial visualization, statistical evaluation, and validation, designed for the urban and rural contexts of Nigeria. All data are from 2020 to ensure consistency, with methods adapted from established frameworks \cite{Parajuli2020}. 

Data sources include case rates from the \texttt{Nigeria Center for Disease Control (NCDC),}poverty from the \texttt{Relative Wealth Index (RWI}), population density from \texttt{WorldPop Hub,} and healthcare access from the \texttt{Humanitarian Data Exchange}. 
The risk score integrates population density (weight 0.2), poverty (0.4), access to healthcare (0.3), and age risk (0.1), adjusted for normalized case rates \cite{Taye2021}. GIS mapping, using \texttt{Python’s GeoPandas}, categorizes states into risk levels to identify hotspots \cite{Taye2021}. Statistical methods, including Spearman correlation, OLS regression, and sensitivity analysis, employ Python libraries \texttt{SciPy, Pandas}, and \texttt{Statsmodels} to ensure reliability \cite{Saltelli2008} \cite{Chatterjee2012}. 

\subsection{Data Sources}
Data were collected from multiple sources to study disease spread:
\begin{enumerate}[label=\roman*]
    \item The COVID-19 case counts from the \texttt{Nigeria Center for Disease Control (NCDC)}, noting possible rural underreporting \cite{Hassan2020COVID19}. 
    \item Population density from \texttt{WorldPop Hub} \cite{WorldPop2020}.
    \item Age structure data from the \texttt{Nigerian Population Commission}, focusing on the proportion of people aged 60 years and older \cite{NPC2020}.
    \item Measurements of poverty from the \texttt{relative wealth index (RWI}) \cite{HDX2020_Wealth}.
    \item Healthcare access from the \texttt{Nigerian Health Facility Registry}, focusing on public facilities \cite{NHFR2020}. 
    \item All data covered the 2020 calendar year to ensure consistency across sources.
\end{enumerate}

\subsection{Data Preparation}
Extensive pre-processing was required to integrate these heterogeneous data sets.

\begin{enumerate}[label=\roman*]
    \item Standardizing regional names across datasets, e.g., \texttt{"Federal Capital Territory"} and \texttt{"FCT"}. All textual data were converted to lowercase for consistency.
    
    \item Aggregating data from \texttt{774} Local Government Areas to \texttt{37}  state-level units (spatial alignment).

    \item No significant missingness was observed across the datasets (e.g., NCDC case counts, WorldPop density, RWI poverty, and health facility registry).

    \item Aligning all data sets to the 2020 calendar year (temporal alignment).
    
    \item Grouping states into low, medium, and high population density categories according to terciles \cite{Taye2021}.
\end{enumerate}

\subsection{Exploratory Data Analysis}
EDA examined the COVID-19 patterns in Nigerian states to inform the composite risk score, using these methods:
\begin{enumerate}[label=\roman*]
    \item \textbf{Temporal analysis} studied weekly case trends in 2020 using Python's Pandas and Seaborn libraries, applying a 7-day rolling average to smooth fluctuations \cite{NCDC2020}. 
    \item \textbf{Spatial analysis} mapped population and density data to compare case distributions between states, examining case rates per 100,000 in low, medium and high density categories to understand the role of density in disease spreads \cite{WorldPop2020}. 
\end{enumerate}

These methods provided a context for vulnerability factors that guided the development of risk scores \cite{Parajuli2020}.

\subsection{Composite Risk Score Construction}
This study develops a composite risk score to prioritize resource allocation by combining structural factors (population density, poverty, healthcare access, age risk) with current COVID-19 case rates. The score identifies states where active outbreaks worsen vulnerabilities, guiding urgent investments in testing, vaccines, or healthcare, rather than simply describing current cases or predicting future ones.

The COVIRA framework, developed in 2020 by Nepalese researchers, is an open source web-based tool (built with PHP, R, and PostgreSQL) designed to assess and communicate the risks of COVID-19 at the individual and regional levels \cite{Parajuli2020}.  It calculates a risk score from 0 to 100 by integrating transmission risk (cases and immigration data), public health risk (underlying conditions and healthcare facilities per population), socioeconomic conditions (poverty, literacy, sanitation), population density and exposure (e.g., airports, occupations) \cite{Parajuli2020}.

The multidisciplinary set-up of COVIRA emphasizes risk communication through visualizations and mechanisms such as public dashboards, making it effective for low-resource settings with limited testing \cite{Parajuli2020}. Its relevance to Nigeria lies in shared challenges, such as lower-middle-income status, rural-urban disparities in healthcare access, and underreporting in remote areas, allowing us to adapt its structure while tailoring weights to the context of Nigeria - for example, greater emphasis on poverty due to its role in rural vulnerabilities \cite{Hassan2020COVID19}.

Case rates per 100,000 are included to adjust the score, ensuring that states with active outbreaks and structural risks are highlighted for urgent resource investment, rather than solely reflecting current case counts.
The composite risk score formula is calculated as:
\begin{equation}
\begin{split}
\text{Risk Score} &= (\alpha \cdot \text{Density} + \beta \cdot \text{Poverty} \\
&\quad + \gamma \cdot \text{Healthcare} + \delta \cdot \text{Age}) \\
&\quad \times \text{Cases per 100k}_{\text{norm}}
\end{split}
\end{equation}

Where $\alpha = 0.2$, $\beta = 0.4$, $\gamma = 0.3$, and $\delta = 0.1$ represent the relative weights of each factor, with $\alpha + \beta + \gamma + \delta = 1.0$.\\

The weights reflect the importance of each factor in Nigeria's vulnerability to COVID-19, informed by the COVIRA framework \cite{Parajuli2020} and the findings of the study. 
\begin{itemize}
    \item \textbf{Poverty} ($\beta = 0.4$, inverted data from the relative wealth index \cite{NHFR2020}) has the highest weight due to its role in limiting healthcare and safe living, especially in rural areas with low reporting \cite{Hassan2020COVID19}\cite{Adesola2024PopulationGrowth}.
    \item Access \textbf{to healthcare} ($\gamma = 0.3$, coverage of public health facilities \cite{Adesola2024PopulationGrowth}) reflects uneven distribution of facilities \cite{NHFR2020}.
    \item \textbf{Population density} ($\alpha = 0.2$, population per square kilometer, classified by terciles \cite{Taye2021}) drives urban cases like Lagos but less in rural areas \cite{Adesola2024PopulationGrowth}.
    \item \textbf{Age risk} ($\delta = 0.1$, proportion of population aged 60 and older \cite{Adesola2024PopulationGrowth}\cite{NPC2020}) is lowest due to Nigeria's young demographic reducing severe outcomes.
\end{itemize}

Factors are normalized to a 0–1 scale using Min-Max scaling:
\begin{equation}
\text{Normalized Score} = \frac{\text{Raw Score} - \text{Min Score}}{\text{Max Score} - \text{Min Score}}
\end{equation}

Normalized cases per 100,000 adjust the score to highlight states with active outbreaks and structural risks, ensuring areas that need urgent resources are highlighted \cite{Parajuli2020}. Sensitivity analysis confirmed stable state rankings in weight adjustments, validating the scoring approach \cite{Saltelli2008}.

\subsection{GIS Mapping}
Spatial mapping visualizes COVID-19 vulnerability in Nigerian states to identify areas that require prioritization of resources. 
\begin{itemize}
    \item \texttt{Python’s GeoPandas} handled spatial data, and \texttt{Matplotlib} created visualizations.
    \item  The state shapefiles, initially in \texttt{WGS84}, were adjusted to \texttt{UTM} coordinates for accurate area calculations \cite{HERA2020}. 
    \item Five choropleth maps showed the risk score, population density, poverty, access to healthcare, and age risk, using different color schemes to differentiate factors \cite{Taye2021}. 
    \item The risk scores were grouped into low, medium and high categories according to percentiles for clear interpretation. 
    \item Maps included state labels, legends, and custom elements for clarity. 
    \item Data were sourced from Humanitarian Data Exchange, with source credits included \cite{HERA2020}. 

\end{itemize}
These maps supported the identification of vulnerability patterns, helping to analyze risk scores.

\subsection{Statistical Analysis}
Statistical analysis supports the development of the composite risk score by examining the relationships between risk factors and the COVID-19 case rates, helping to prioritize resource allocation.
\begin{itemize}
    \item The Spearman rank correlation analyzed pairwise relationships between factors:\textbf{ population density, poverty, access to healthcare}, and \textbf{age risk}, selected for its compatibility with data not normally distributed \cite{Saltelli2008}. 
    \item Ordinary Least Squares regression, conducted with \texttt{Statsmodels}, used normalized case rates per 100,000 as the outcome.
    \item  The four factors as predictors, defined as:
\begin{equation}
        \text{Cases per 100k}_{\text{norm}} = \frac{C - C{\text{min}}}{C_{\text{max}} - C_{\text{min}}}
    \end{equation}
    
    Where $C$ represents the cases per 100,000 population, and $C_{\text{min}}$ and $C_{\text{max}}$ represent the minimum and maximum values across all states, respectively. 
    
    \item The sensitivity analysis adjusted the factor weights, keeping their sum at \texttt{1.0}, to test the stability of the risk score by comparing the rankings \cite{Saltelli2008}. 
    \item Visualizations, created with \texttt{Seaborn}, explored factor distributions and relationships [8]. 
    \item The \texttt{SciPy} and \texttt{Statsmodels} Python libraries supported these analyzes, providing a basis for validating the risk score \cite{Chatterjee2012}.
\end{itemize}

\subsection{Validation and Robustness}
Validation and robustness checks ensure that the composite risk score reliably identifies states for resource prioritization.
\begin{itemize}
    \item Cross-validation compared risk scores with Nigeria Centre for Disease Control (NCDC) epidemiological reports to verify alignment with observed patterns \cite{NCDC2020}. 
    \item A collinearity test, using variance inflation factor analysis, evaluated factor overlap to inform regression interpretation \cite{Saltelli2008}. 
    \item Stability testing adjusted factor weights and tested alternative models, such as logarithmic case rates, to confirm consistent state rankings \cite{Saltelli2008}. 
    \item The outlier analysis reviewed extreme data points, retaining valid cases such as high-density states for their epidemiological importance \cite{Taye2021}. 
\end{itemize}
Limitations included potential rural underreporting, reliance on facility counts over quality, use of static 2020 data, and lack of mobility data.
These procedures validated the methodology, ensuring an accurate vulnerability assessment while acknowledging data constraints \cite{Saltelli2008}.

\section{Results}
This section presents findings on COVID-19 vulnerability in Nigerian states using 2020 data from the Nigeria Centre for Disease Control (NCDC) and other sources \cite{NCDC2020}. The results focus on key areas: \textbf{patterns over time, regional differences}, and \textbf{statistical relationships}, aiming to identify high-risk areas and inform public health strategies. 

\subsection{Patterns Over Time}
\begin{figure}[htbp]
\centering
\begin{subfigure}[b]{0.48\columnwidth}
    \centering
    \includegraphics[width=\columnwidth]{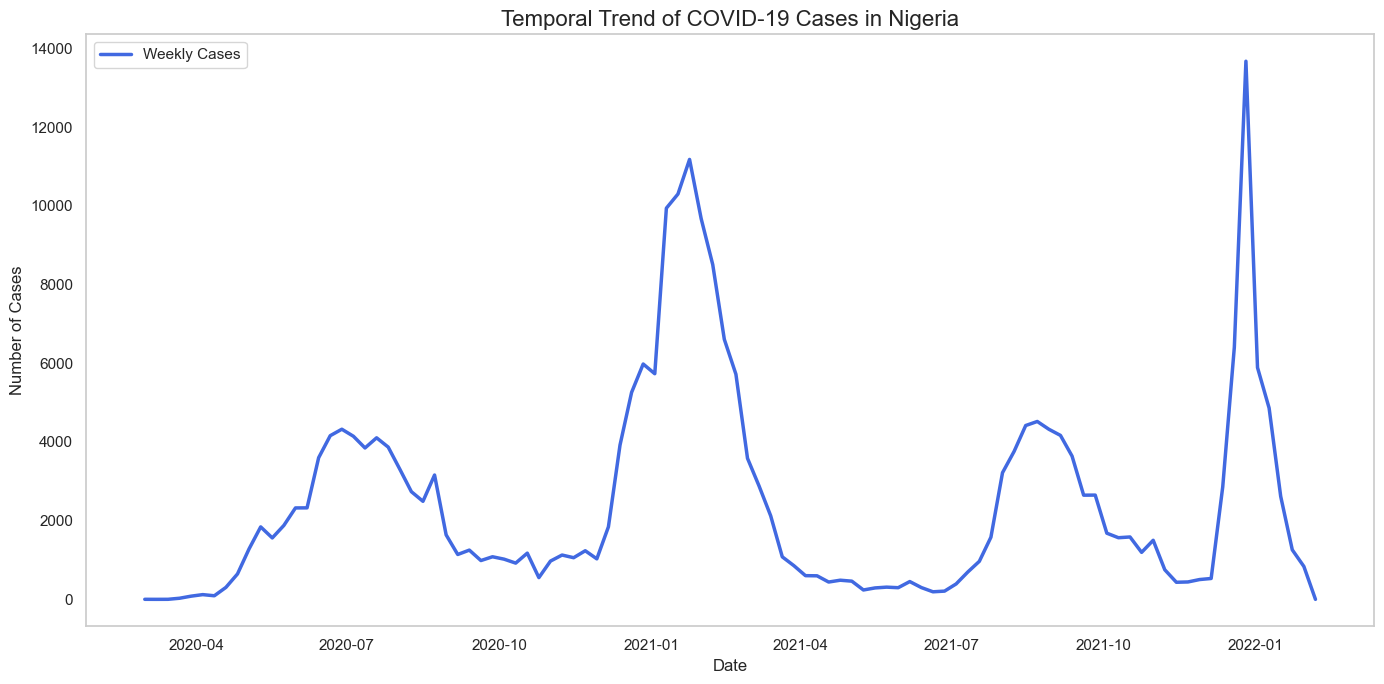}
    \caption{Temporal trend of COVID-19 cases}
    \label{fig:temporal_trend}
\end{subfigure}
\hfill
\begin{subfigure}[b]{0.48\columnwidth}
    \centering
    \includegraphics[width=\columnwidth]{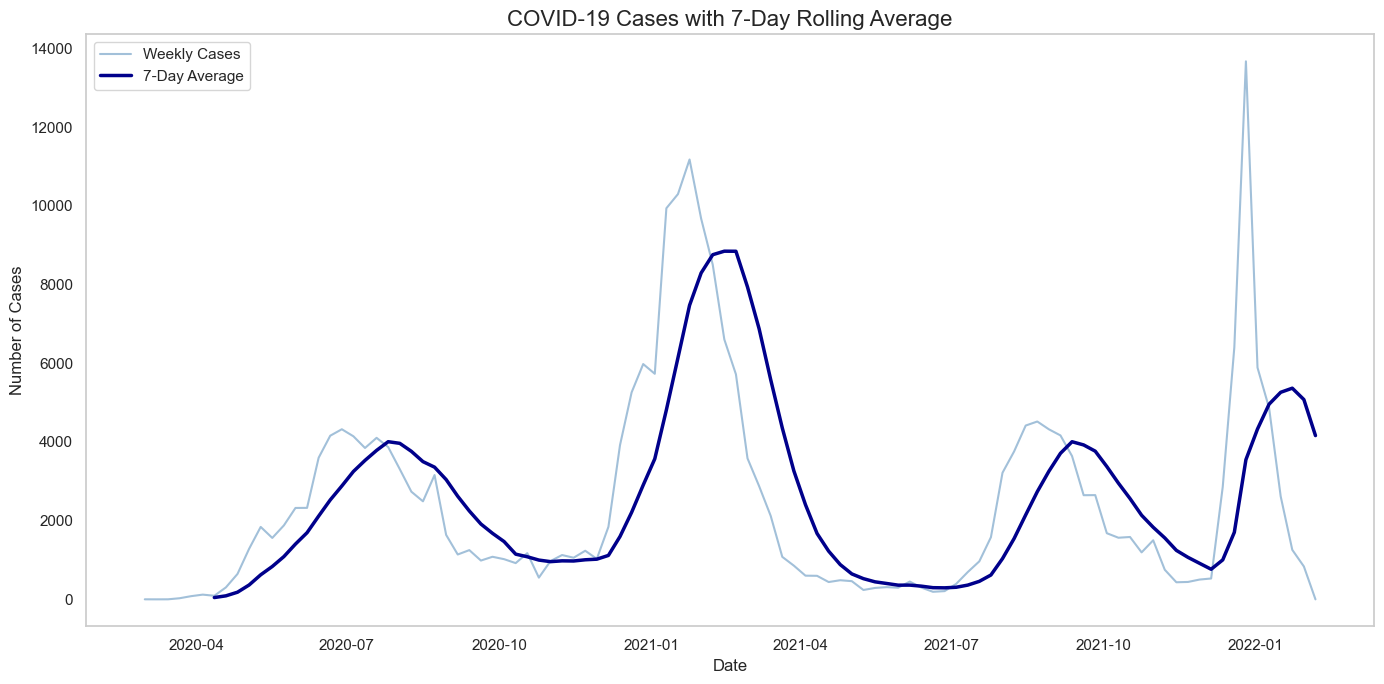}
    \caption{Cases with 7-day rolling average}
    \label{fig:rolling_average}
\end{subfigure}

\vspace{0.5cm}

\begin{subfigure}[b]{\columnwidth}
    \centering
    \includegraphics[width=\columnwidth]{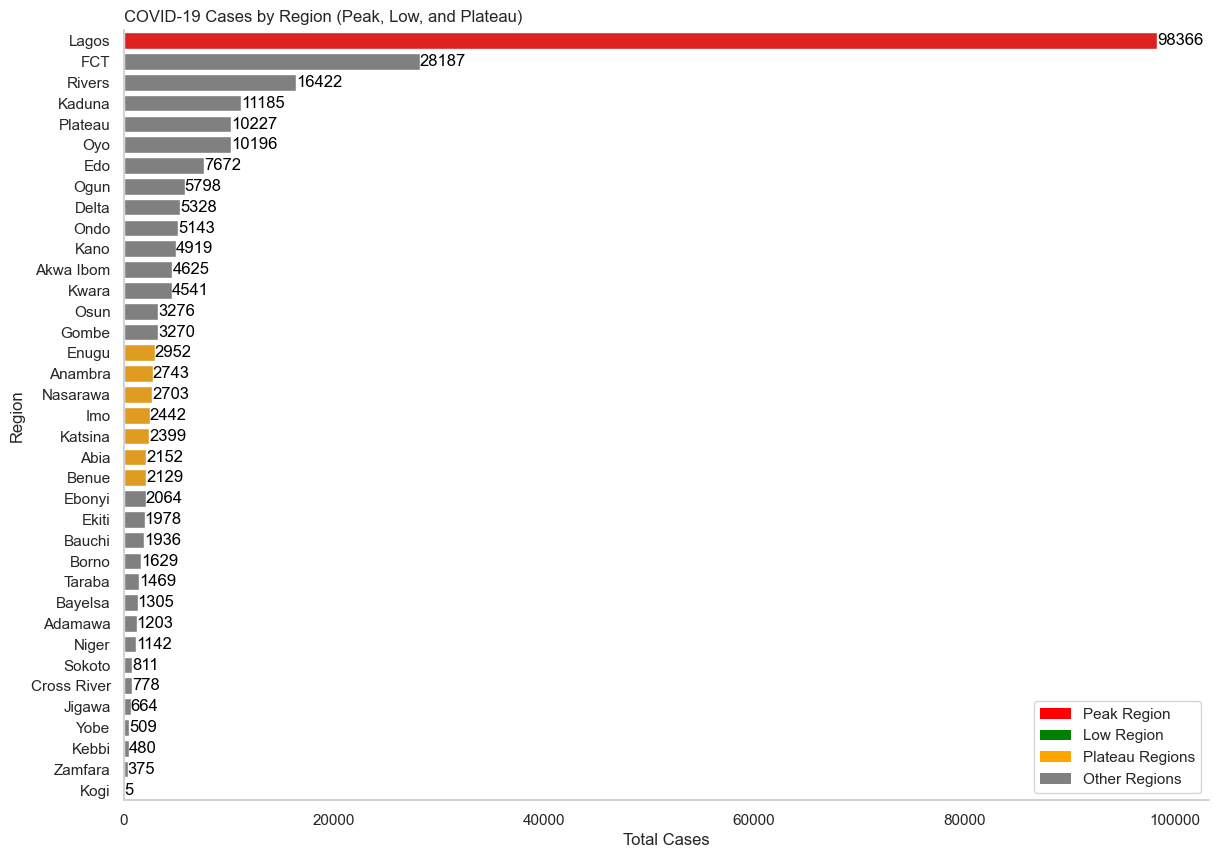}
    \caption{COVID-19 cases by region}
    \label{fig:covid_cases_by_region}
\end{subfigure}

\caption{Temporal patterns of COVID-19 in Nigeria: (a) weekly case trends showing four distinct infection waves, (b) smoothed trends using 7-day rolling average, and (c) regional distribution with Lagos as the peak region}
\label{fig:temporal_combined}
\end{figure}

Temporal analysis revealed four distinct COVID-19 waves in Nigeria (Fig. \ref {fig:temporal_trend}): July 2020 (1,000 cases/week), January-March 2021 (2,000 cases/week), mid-2021 (1,500 cases/week), and December 2021 (3,000 cases/week). These waves reflected the ease of lockdown, new variants, and a decreased adherence to safety measures \cite{NCDC2020}. 

The regional distribution of cases (Fig. \ref{fig:covid_cases_by_region}) varied significantly between the 37 states of Nigeria and the FCT. Lagos recorded the highest burden (98,366 cases) due to its population density and transport hub status, while states such as Kogi reported minimal cases (5 cases), likely reflecting limited testing or geographic isolation. This urban-rural disparity demonstrates the uneven burden of diseases in Nigeria. \cite{NCDC2020}.

Google Trends analysis revealed a decrease in public interest in COVID-19 information after March 2020, with a weak correlation to actual case rates (\texttt{r = 0.0415}), suggesting a limited utility of online search data for tracking the spread of the disease in Nigeria \cite{Nuti2014GoogleTrends}.

\subsection{Regional Patterns}
\begin{figure}[htbp]
\centering
\begin{subfigure}[b]{0.48\columnwidth}
    \centering
    \includegraphics[width=\columnwidth]{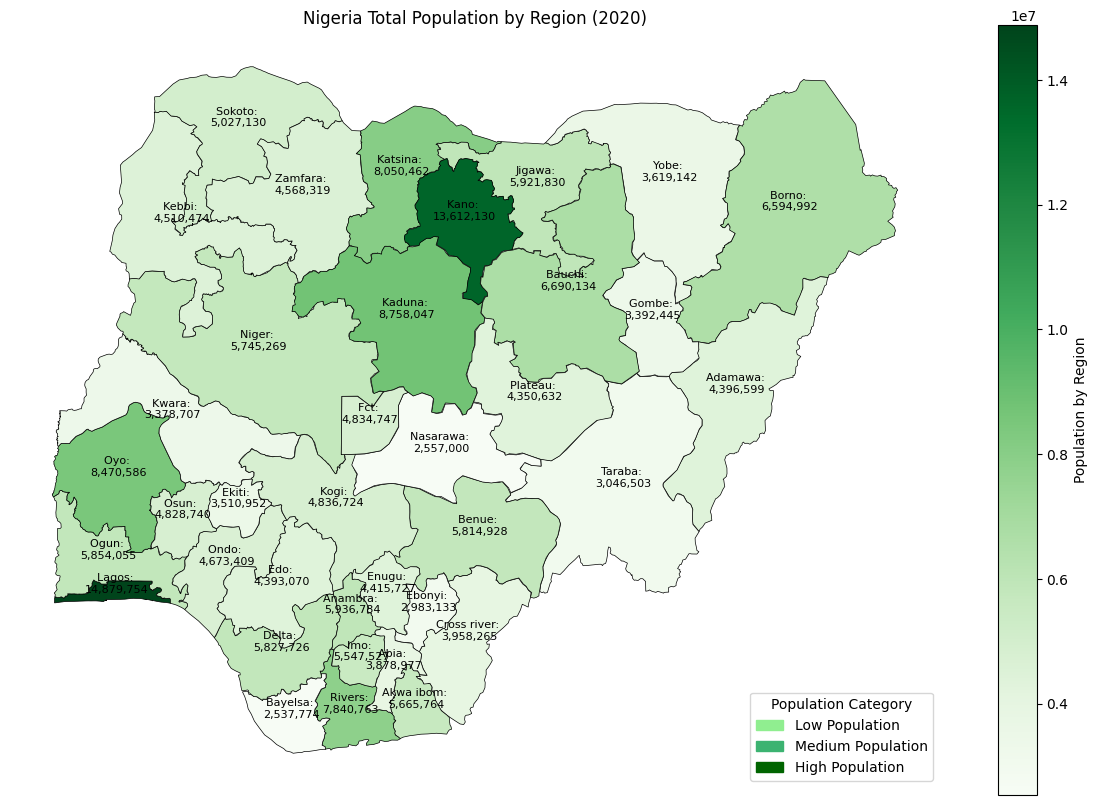}
    \caption{Nigeria Population By Region}
    \label{fig:population_count}
\end{subfigure}
\hfill
\begin{subfigure}[b]{0.48\columnwidth}
    \centering
    \includegraphics[width=\columnwidth]{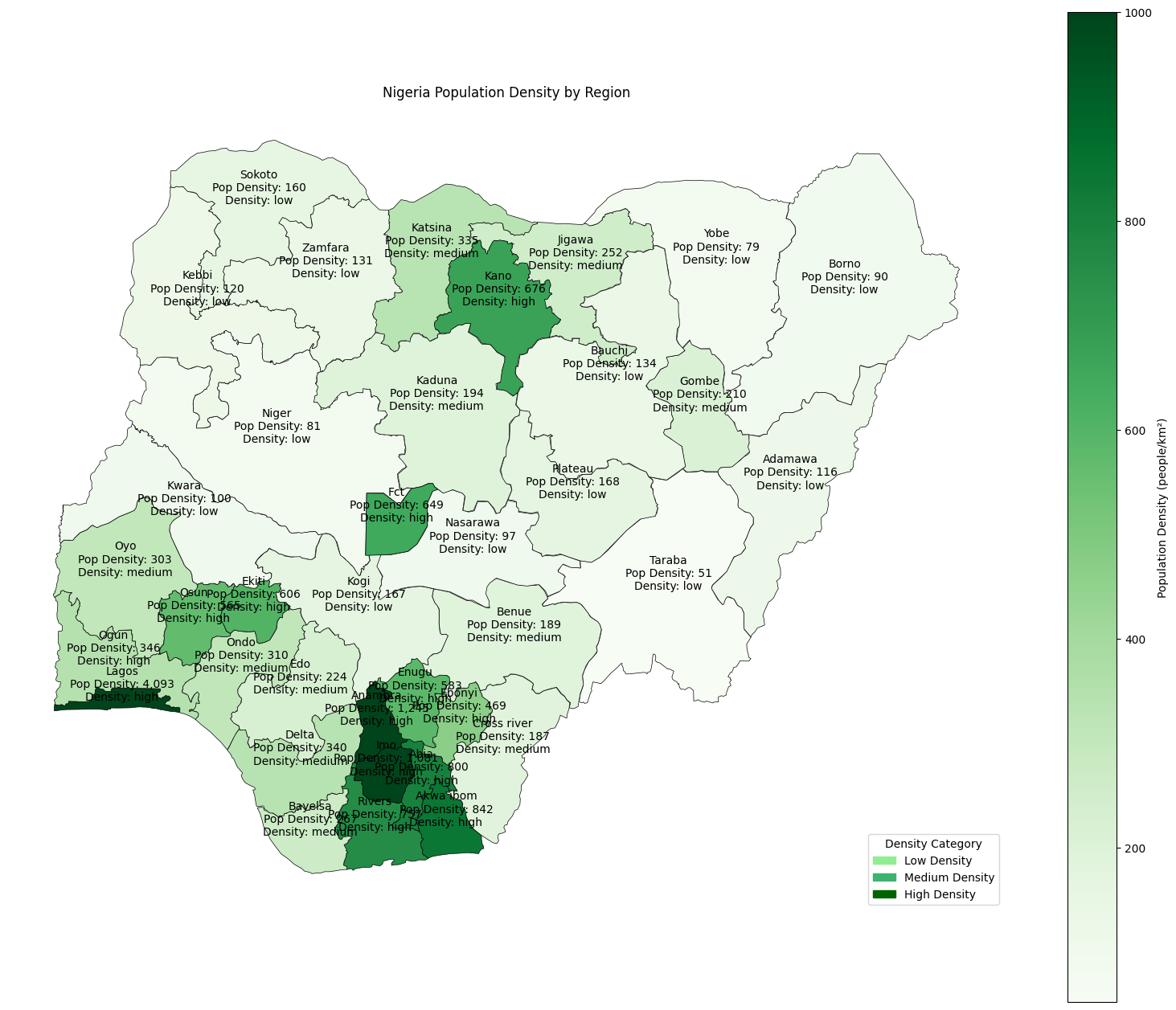}
    \caption{Nigeria Population Density By Region}
    \label{fig:population_density}
\end{subfigure}
\caption{Population distribution and density across Nigerian states}
\label{fig:population_combined}
\end{figure}

Spatial analysis explored how population and density influenced the distribution of the cases. Fig. \ref{fig:population_count}, a population distribution map, used a \texttt{green-to-black} gradient to show that Lagos and FCT had the largest populations, while northern states such as Bauchi and Yobe had smaller populations \cite{WorldPop2020}. Fig. \ref{fig:population_density}, a density map, used a \texttt{light-to-dark green} gradient to highlight Lagos (\texttt{7,777 people/km²}) and FCT as the densest areas, compared to low-density states such as Bauchi and Yobe \cite{WorldPop2020}.

\begin{figure}[h!]
    \centering
    \includegraphics[width=0.4\textwidth]{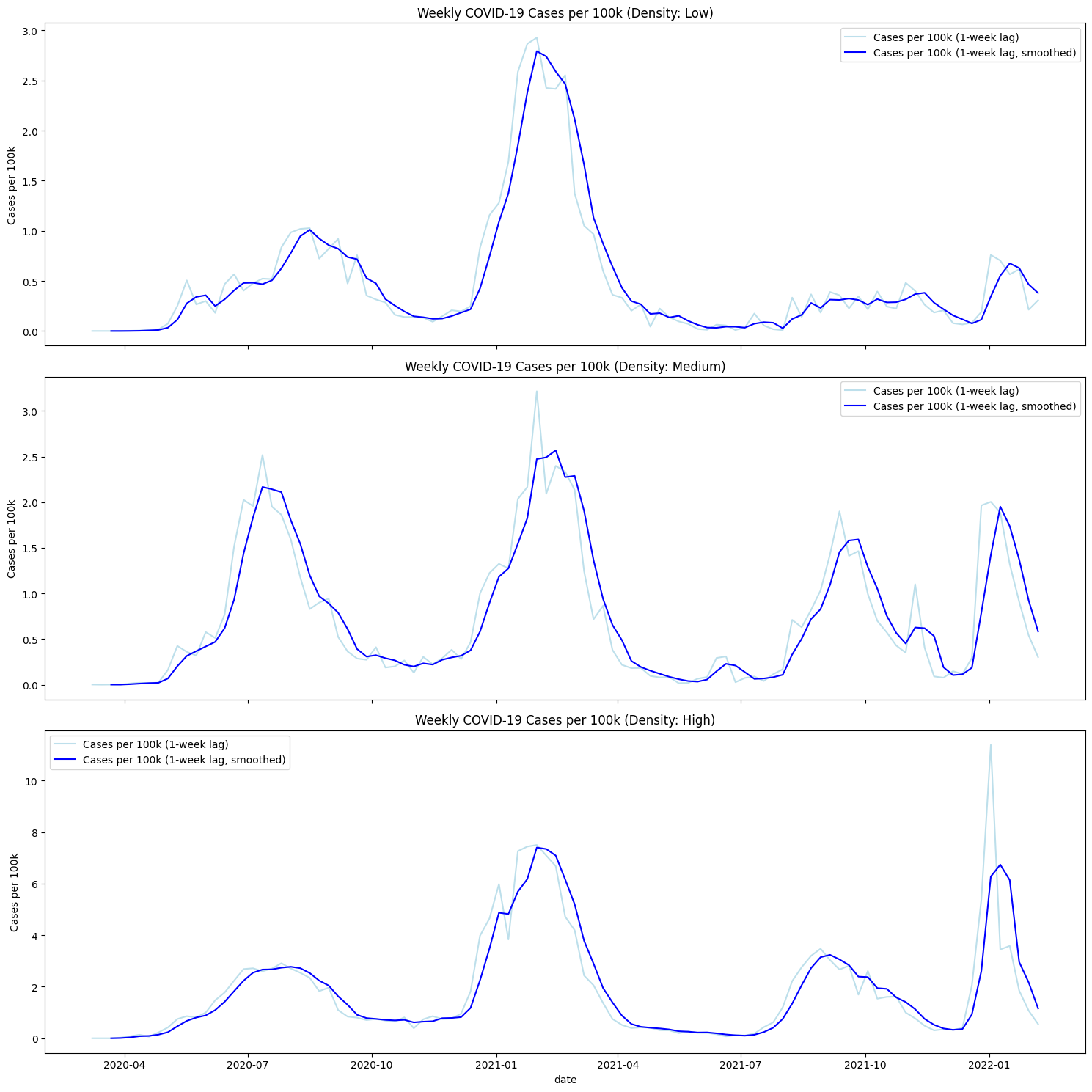}
    \caption{\small Weekly COVID-19 Cases per 100,000 (Smoothed vs. Non-Smoothed Across The Three Groups}
    \label{fig:smoothed_weekly_covid_cases}
\end{figure}

The case rates per 100,000 people were then compared across states grouped by density (low, medium, high). 
\[
\text{Cases per 100K} = \left( \frac{\text{Number of Cases}}{\text{Population}} \right) \times 100,000
\]
Fig. \ref{fig:smoothed_weekly_covid_cases}  presents three line graphs that track these rates from April 2020 to January 2022. 

The low-density states peaked at \texttt{2.8 cases per 100,000} in January 2021, the medium-density states at \texttt{2.5}, and high-density states at \texttt{10 per 100,000} in January 2022. This pattern indicates that denser areas experienced higher infection rates, likely due to increased opportunities for transmission \cite{WorldPop2020}.       

\subsection{Risk Mapping}
GIS mapping visualized vulnerability through a composite risk score and its contributing factors. Fig. \ref{fig:covid-risk-factors}, the risk score map, in \texttt{pastel} shades, identifies high-risk states. 

Lagos topped the list, driven by its density and case load, while northern states such as Sokoto and Zamfara showed an increased risk due to poverty and limited access to healthcare despite fewer cases \cite{HERA2020}. 

Individual factor maps further detail these drivers, showing urban density and rural socioeconomic challenges as key vulnerability factors \cite{HERA2020}:

\begin{figure}[htbp]
\centering
\begin{subfigure}[b]{0.48\columnwidth}
    \centering
    \includegraphics[width=\linewidth]{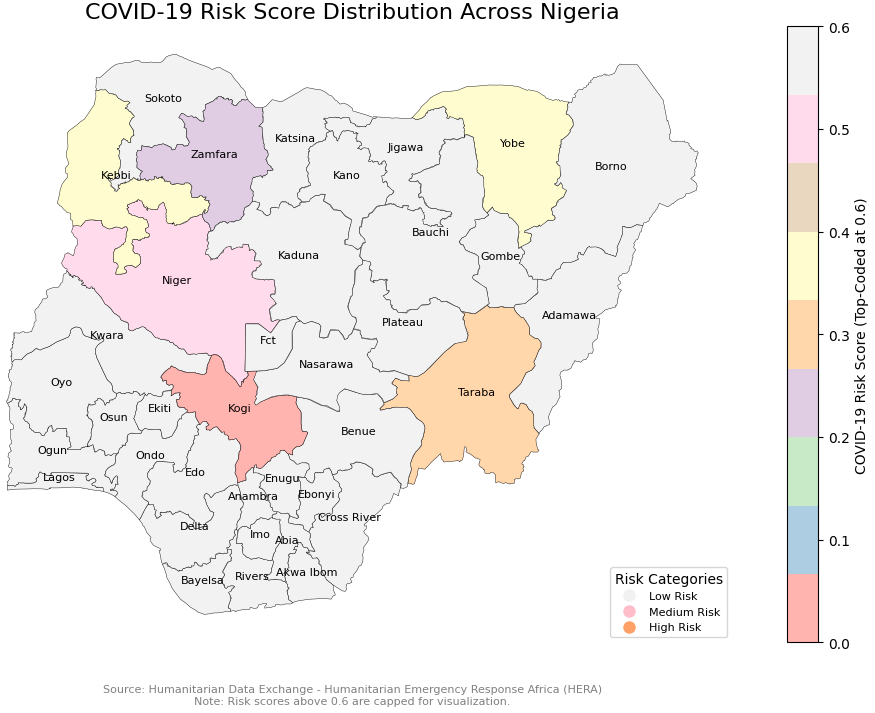}
    \caption{COVID-19 Risk Factors Across States}
    \label{fig:covid-risk-factors}
\end{subfigure}
\hfill
\begin{subfigure}[b]{0.48\columnwidth}
    \centering
    \includegraphics[width=\linewidth]{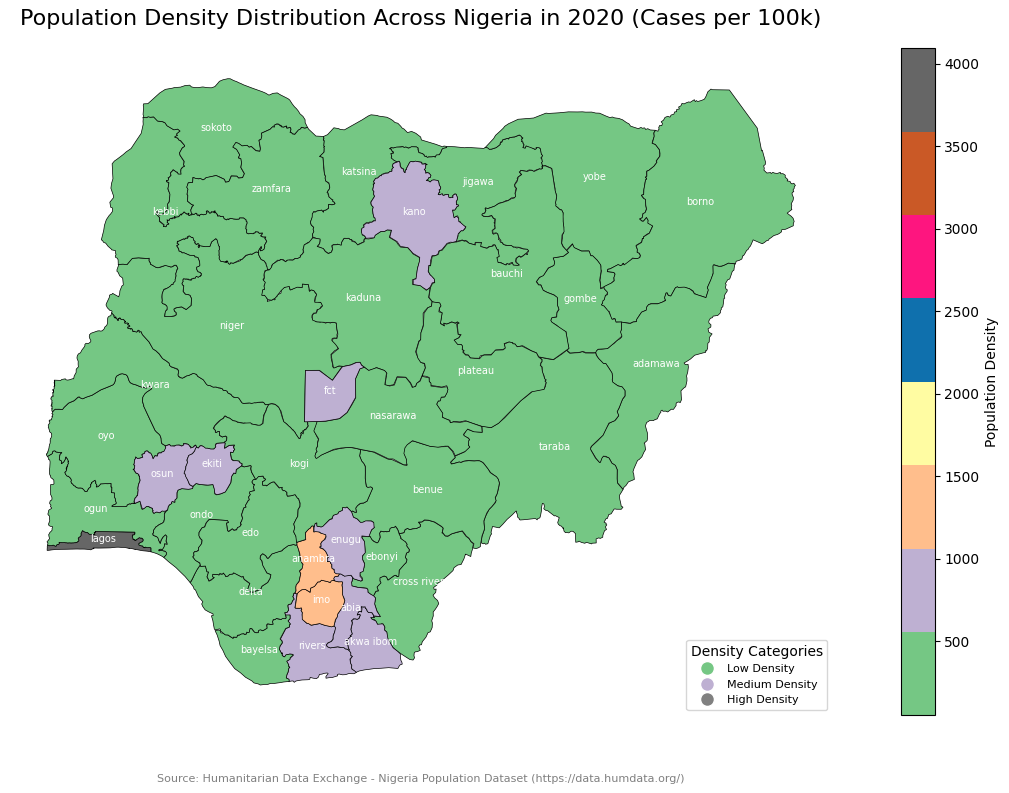}
    \caption{Population Density Distribution}
    \label{fig:green-shades-pop-density}
\end{subfigure}

\vspace{0.3cm}

\begin{subfigure}[b]{0.48\columnwidth}
    \centering
    \includegraphics[width=\linewidth]{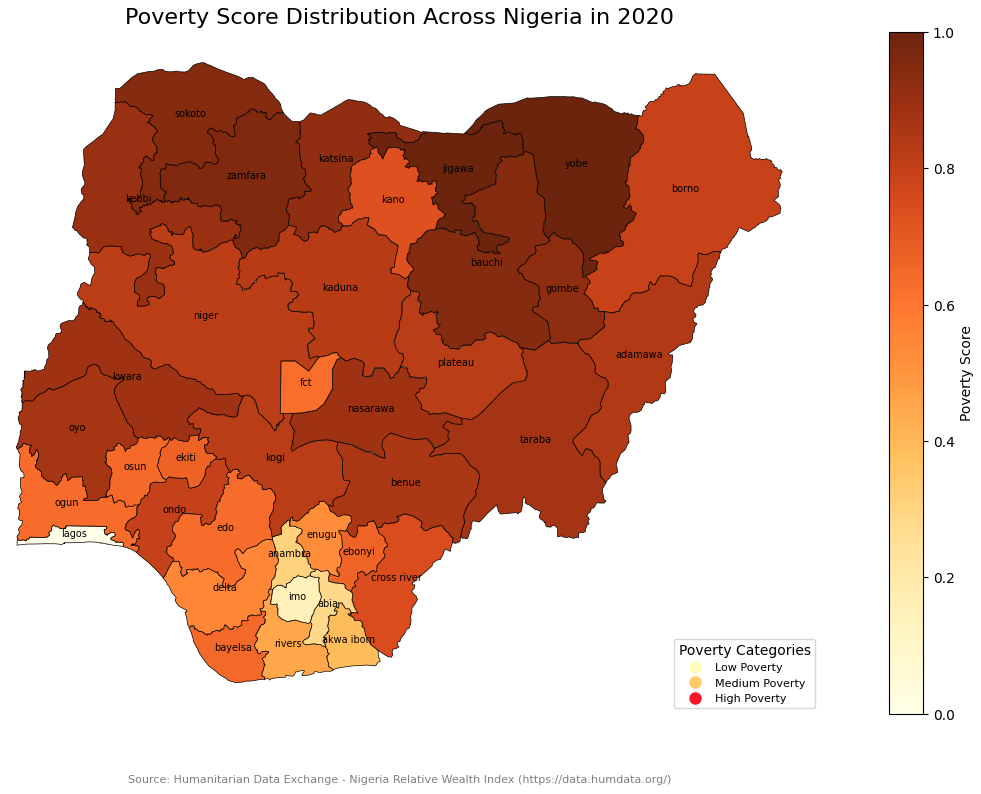}
    \caption{Poverty Score Distribution}
    \label{fig:poverty-map}
\end{subfigure}
\hfill
\begin{subfigure}[b]{0.48\columnwidth}
    \centering
    \includegraphics[width=\linewidth]{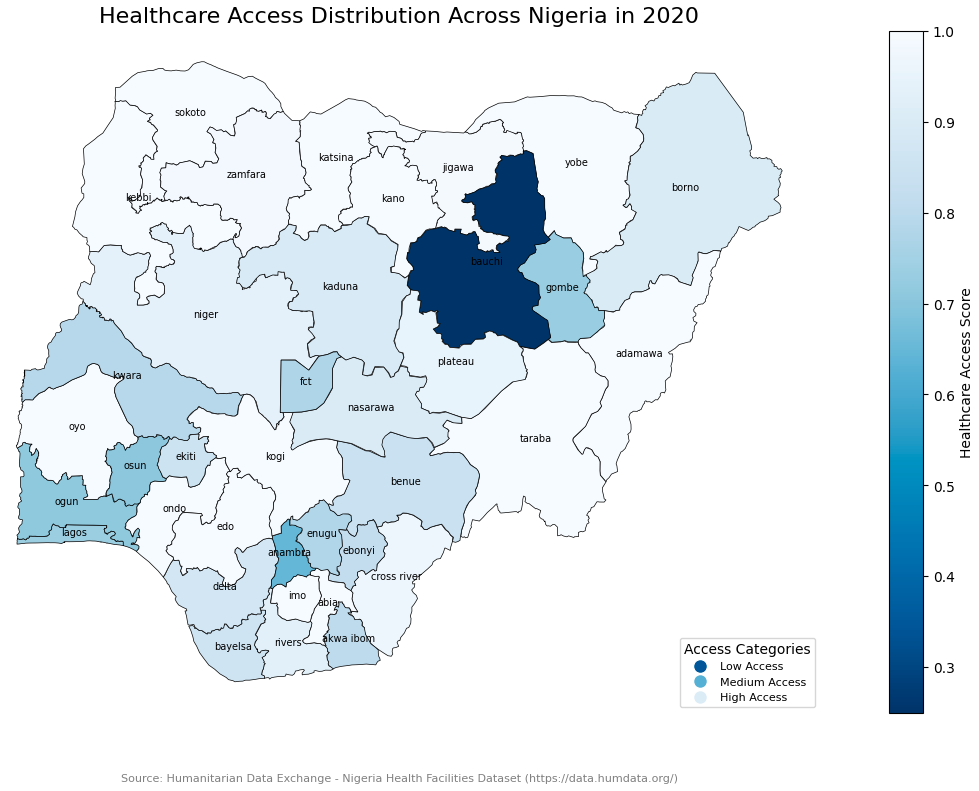}
    \caption{Healthcare Access Distribution}
    \label{fig:healthcare-access}
\end{subfigure}

\vspace{0.3cm}

\begin{subfigure}[b]{0.48\columnwidth}
    \centering
    \includegraphics[width=\linewidth]{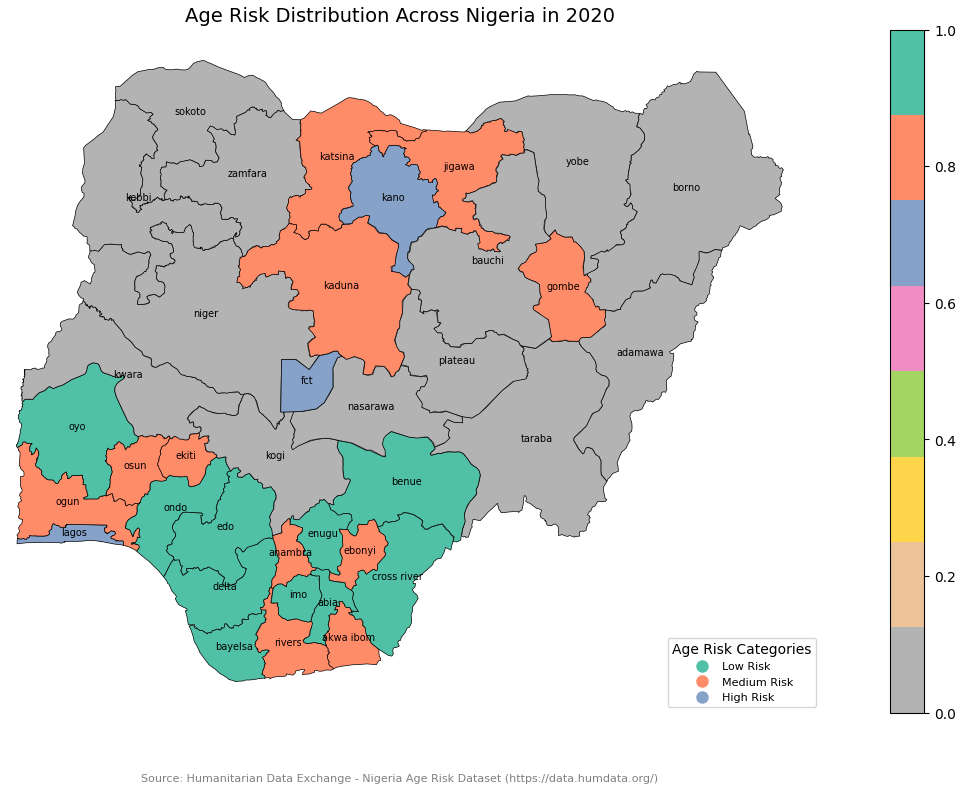}
    \caption{Age Risk Distribution Map}
    \label{fig:age-risk-map}
\end{subfigure}
\caption{Risk factor maps showing spatial distribution of COVID-19 vulnerability components.}
\label{fig:risk_factors_combined}
\end{figure}

\subsection{Statistical Findings}
Statistical analysis examined relationships between risk factors and case rates. \texttt{Spearman’s correlation} revealed the following:

\subsubsection{Correlation Analysis}
\begin{enumerate}
    \item  \textbf{Moderate positive} link between \texttt{density} and \texttt{cases} (\texttt{r = 0.37, p < 0.05}).
    \item \textbf{Strong negative} link between \texttt{poverty} and \texttt{density} (\texttt{r = -0.77, p < 0.01}). 
    \item \textbf{Weaker correlations} for \texttt{healthcare access} (\texttt{r = -0.31, p < 0.05}), and \texttt{age risk} (\texttt{r = 0.26, p < 0.05}).
\end{enumerate}

\subsubsection{Regression Analysis}
\begin{figure}[h!]
    \centering
    \includegraphics[width=0.4\textwidth]{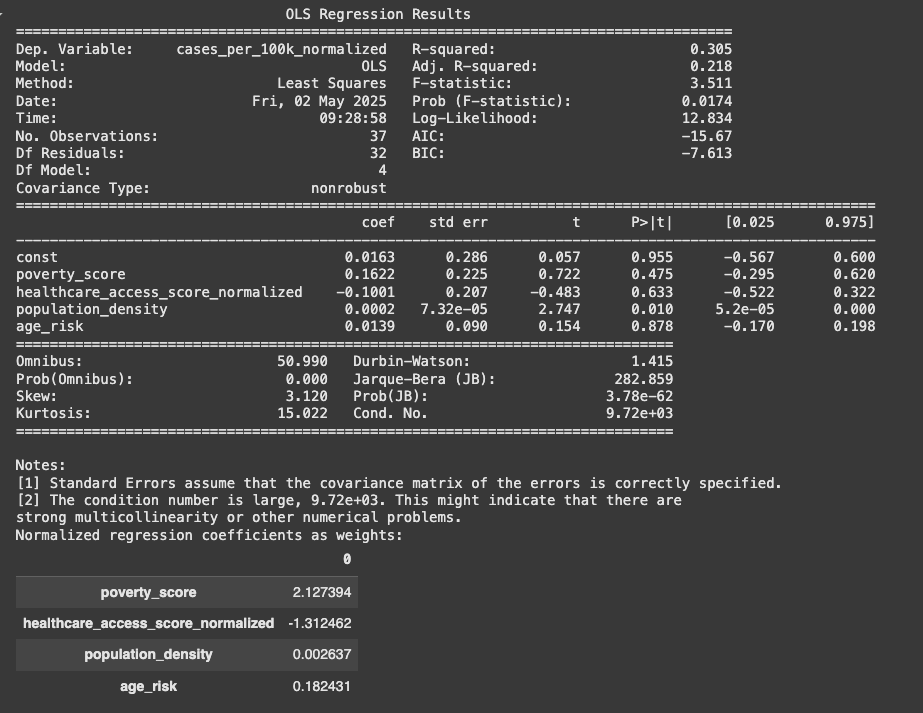}
    \caption{\small Regression Analysis}
    \label{fig:regression-ols-analysis}
\end{figure}

The regression model as shown in Fig. \ref{fig:regression-ols-analysis} explained \texttt{30.5\%} of the differences in case rates between states ($R^2 = 0.305$). This suggests that the model captures a moderate amount of what drives infection rates, although other factors not included in the model account for the remaining \texttt{69.5\%}\cite{Saltelli2008}. 
 
 The model was validated using standard regression diagnostics, with no train-test split applied, as the analysis focuses on describing current vulnerability rather than prediction.

However, a high condition number (\texttt{9.72e+03}) suggested some overlap among the factors, which could reduce the precision of the results \cite{Saltelli2008}.

\subsubsection{Sensitivity Analysis}
Adjusted poverty weights (\texttt{0.3} to \texttt{0.5}), did not change state rankings, confirming model robustness.  

\begin{figure}[htbp]
\centering
\begin{subfigure}[b]{0.48\columnwidth}
    \centering
    \includegraphics[width=\columnwidth]{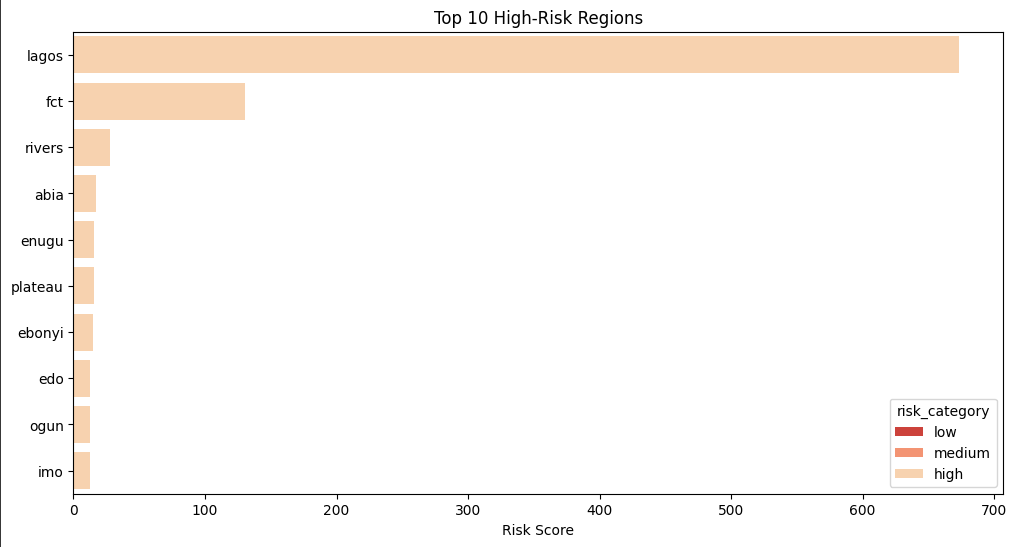}
    \caption{Top 10 High Risk Regions}
    \label{fig:top-10-high-risk-regions}
\end{subfigure}
\hfill
\begin{subfigure}[b]{0.48\columnwidth}
    \centering
    \includegraphics[width=\columnwidth]{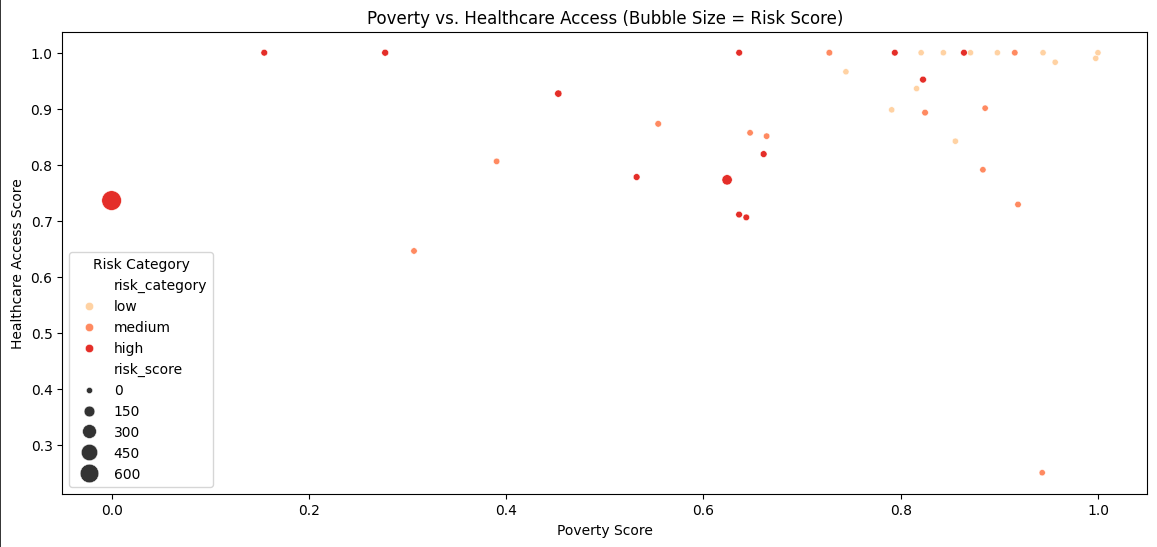}
    \caption{Poverty vs Healthcare}
    \label{fig:poverty-vs-healthcare}
\end{subfigure}

\vspace{0.3cm}

\begin{subfigure}[b]{0.48\columnwidth}
    \centering
    \includegraphics[width=\columnwidth]{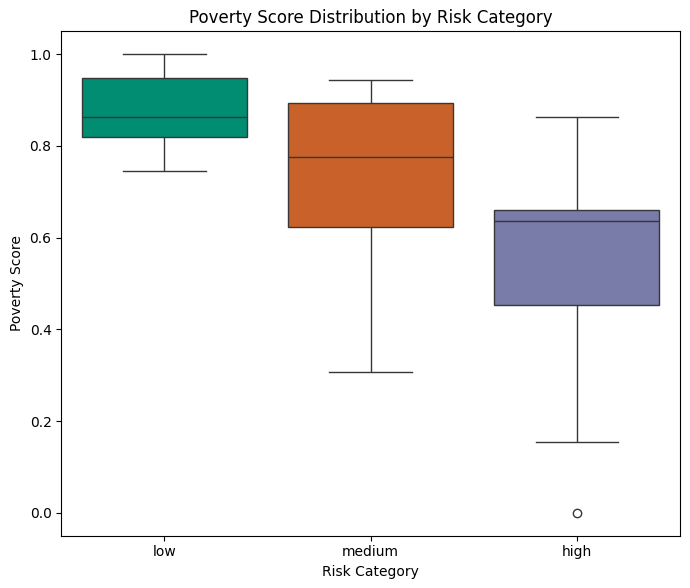}
    \caption{Poverty Score Box Plot}
    \label{fig:poverty-box-plot}
\end{subfigure}
\hfill
\begin{subfigure}[b]{0.48\columnwidth}
    \centering
    \includegraphics[width=\columnwidth]{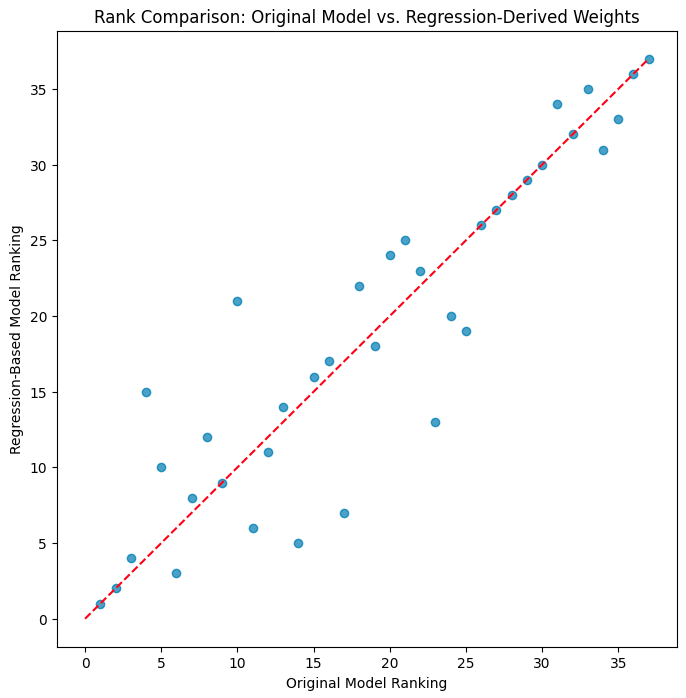}
    \caption{Rank Comparison}
    \label{fig:rank-comparison}
\end{subfigure}

\caption{Statistical visualizations supporting COVID-19 vulnerability analysis}
\label{fig:statistical_combined}
\end{figure}

Statistical visualizations (Fig. 6) validated these findings:
\begin{enumerate}
    \item  The bar graph in Fig. \ref{fig:top-10-high-risk-regions} identified Lagos and the Federal Capital Territory (FCT) as areas of highest risk.
    \item  The scatter plot in Fig. \ref{fig:poverty-vs-healthcare}, revealed poverty-health care relationships between states.
    \item  The box plot in Fig. \ref{fig:poverty-box-plot} confirmed the risk factor distributions by category \cite{Saltelli2008}.
    \item The ranking comparisons in Fig. \ref{fig:rank-comparison} demonstrated consistent state classifications regardless of weighting method, proving the robustness of the model \cite{Chatterjee2012}.
\end{enumerate}

\subsection{Validation of Results}
The risk score was validated against NCDC reports, with high-risk areas such as Lagos and FCT aligned with known hotspots, confirming the model precision. 

Factor overlap analysis revealed significant poverty-density correlations, which explains the multicollinearity in the regression analysis. Stability tests consistently identified the same high-risk states despite weight adjustments, demonstrating model robustness \cite{Saltelli2008} . Lagos was retained despite extreme density due to epidemiological significance \cite{Taye2021}.

Several limitations were noted:
\begin{enumerate}
    \item Potential underreporting in rural areas that affects case rate normalization \cite{NCDC2020}. 
    \item Healthcare data based on facility counts rather than quality/capacity. 
    \item Static 2020 data limit temporal dynamics assessment. 
    \item Excluded mobility data that could overlook transmission factors. 
\end{enumerate}
The results provide a robust vulnerability assessment for COVID-19 to guide public health policies.

\section{Discussion}
This study examines the factors driving the vulnerability of COVID-19 in Nigerian states, using a composite risk score that combines population density, poverty, access to healthcare and age risk, adjusted for case rates. The results confirm that high-density urban areas such as Lagos (risk score 673.47 vs. national average 28.16) had the highest case counts \cite{NCDC2020}, aligning with Taye and Popoola's findings on urban hotspots \cite{Taye2021}. Unlike single-factor studies (e.g. density \cite{Taye2021}), our holistic approach highlights poverty (weight 0.4) and access to healthcare (0.3) as critical, particularly in rural areas with lower density with socioeconomic challenges despite lower density.

Rural underreporting, as noted by Hassan and Hashim \cite{Hassan2020COVID19}, probably underestimates cases due to limited testing and data collection, which means that the actual vulnerability in these areas could be higher than the risk score suggests. This may overemphasize urban impacts, where testing was stronger (e.g. Lagos). However, poverty and access to healthcare remain critical risk factors \cite{Hassan2020COVID19}.

Temporal analysis revealed four infection waves consistent with the NCDC reports, driven by relaxed measures, gatherings, and variants such as Delta \cite{NCDC2020}. Google Trends data, as a supplementary analysis, showed declining public interest after March 2020 (correlation\texttt{ r = 0.0415}), possibly indicating fatigue or alternative sources of information such as radio, as noted by Adesola et al. \cite{Adesola2024PopulationGrowth}. This contextualizes awareness, but is not central to the risk score.

Our methodology adapts the \texttt{COVIRA} framework from Nepal \cite{Parajuli2020}, where population density was a significant predictor ($\beta = 0.42$, $p < 0.01$, which means that density had a strong and statistically significant effect on risk).  In our study, the regression results (Section III.G) found that population density is significant, but our higher poverty weighting (0.4 vs. 0.2 for density) emphasizes the socioeconomic barriers of Nigeria \cite{NCDC2020}. 

Compared to global models such as the CDC's Social Vulnerability Index, which guides the allocation of resources for pandemics \cite{Chien2024}, our score prioritizes the rural-urban disparities of Nigeria. Sensitivity analysis, detailed in Section III.H, adjusted the weights (e.g. poverty from 0.4 to 0.3 or 0.5) and confirmed stable state rankings, ensuring reliability despite moderate correlations such as poverty density (\texttt{r = 0.37, p < 0.05}), from Section III.G.

\section{Implications (Ethical Considerations)}
The risk score serves as a tool for directing health resources to high-risk areas (e.g. Lagos, Abuja) and rural areas with limited healthcare, promoting equitable interventions. However, labeling regions as \texttt{'high risk'} could stigmatize communities, worsening social or economic divides if support is not fair. 

Using data from Google Trends and Relative Wealth Index raises privacy concerns. Although aggregated and anonymized, grouping areas by risk could be misused without transparency. 

Under-reporting in rural areas \cite{Hassan2020COVID19}, can skew risk scores, potentially diverting resources from hidden hotspots. Deploying mobile teams for better testing and data collection could ensure fairer allocation, aligning with the goal of scoring prioritization.

\section{Conclusion}
This study addresses: \textbf{What are the primary factors contributing to COVID-19 vulnerability in Nigerian states, and how can they be quantified using a composite risk score to inform decision making?} 

The findings highlight population density, poverty, access to healthcare, and age risk as key drivers. The risk score, validated against NCDC reports \cite{NCDC2020}, reliably identifies high-risk urban areas such as Lagos and vulnerable rural regions, reflecting the diverse health challenges of Nigeria.

Limitations include reliance on 2020 data, excluding factors such as mobility or vaccinations, and rural under-reporting from weak tests \cite{Hassan2020COVID19}, potentially underestimating rural risks. Future research could use real-time data, include factors, e.g. vaccine coverage, or apply the risk score to other African countries. 

This work provides guidance for health crises in Nigeria, bridging data-driven analysis with policy implications in its complex socio-epidemiological environment. Although focused on COVID-19, the framework can extend to other infectious diseases, offering lessons for future pandemics in resource-limited settings.



\end{document}